\documentclass[amsmat,amssymb,amsfonts,aps,prb,twocolumn,showpacs]{revtex4}

\usepackage{graphicx}
\usepackage{dcolumn}
\usepackage{bm}
\begin{document}

\title{Magnetism in graphene nano-islands}

\author{J. Fern\'andez-Rossier$^1$, J. J. Palacios$^{1,2}$}
\affiliation{ (1)Departamento de F\'{\i}sica Aplicada, Universidad de Alicante, 
San Vicente del Raspeig, Alicante E-03690, Spain. 
\\(2)Instituto de Ciencia de Materiales de Madrid
(CSIC), Cantoblanco, Madrid E-28049, Spain.}

\date{\today} 

\begin{abstract} 
We study the magnetic properties of nanometer-sized graphene structures 
with triangular and  hexagonal shapes  terminated by 
zigzag edges. We discuss how the shape of the island, 
the imbalance in the number of atoms belonging to the two graphene sublattices,
the existence of zero-energy states, and the total and local magnetic moment 
are intimately related. We consider electronic interactions both in
a mean-field approximation of the one-orbital Hubbard model and with density
functional calculations.  Both descriptions yield values for the ground state
total spin,  $S$, consistent with Lieb's theorem for bipartite lattices.
Triangles have a finite  $S$ for all sizes whereas
hexagons have $S=0$ and develop local moments above a critical size 
of $\approx 1.5$ nm.


\end{abstract}

\maketitle


The study of graphene-based field effect devices  has opened a new research
venue in nanoelectronics\cite{Novoselov04,Bunch05,Geim05,Kim05,Science06}.
 Graphene is a truly two-dimensional
zero-gap semiconductor with peculiar transport and magnetotransport  
properties, including room temperature Quantum Hall effect\cite{RoomQHE},
that makes it
very different from conventional semiconductors and metals\cite{Natmat}.
 Progress in the fabrication of graphene-based lower dimensional structures have
been reported both in the form of one-dimensional ribbons\cite{Avouris,Melinda} 
and zero-dimensional dots\cite{Bunch05,Natmat,Pablo07}. Interestingly,
the  electronic properties of graphene change in a non-trivial manner when going
to lower dimensions.  Ribbons, for instance, can be either semiconducting with
a size dependent   gap or metallic\cite{Avouris,Melinda}. 

The electronic structure of graphene-based nanostructures is
expected to  be different
from bulk graphene because of surface, or, more properly,
 edge effects\cite{Nakada96}. This
is particularly true in the case  of structures
with ziz-zag edges which present magnetic properties
\cite{Fujita96,HMzz,Waka98}. 
Whereas bulk graphene is 
a diamagnetic semimetal, simple tight-binding models predict that
one-dimensional ribbons with zigzag edges have two flat bands at the Fermi
energy \cite{Nakada96,Fujita96,Ezawa06,Brey06,Fede06}, i.e., 
are paramagnetic metals. 
Spin polarized density functional theory (DFT)\cite{HMzz} and mean
field\cite{Fujita96} 
 calculations confirm that these bands are prone to magnetic
instabilities. 

  
The fabrication of graphene nanostructures using top-bottom  techniques
does not permit creating atomically defined edges to date\cite{Pablo07}. 
In contrast, bottom-up processing of
graphene nano-islands is very promising \cite{Wu07}. 
 Hexagonal shape nano-islands  with  well-defined zigzag edges atop 
the $0001$ surface of Ru  have already been achieved\cite{Rodolfo}.
This experimental progress motivates our study of the electronic
structure of graphene nanostructures with various shapes. 
Graphene quantum dots also hold the promise of 
extremely long spin relaxation and decoherence time because of  
the  very  small spin-orbit and hyperfine coupling in carbon\cite{Loss}.

We have found that, remarkably, both the DFT calculations and the
mean field approximation of the single-band Hubbard model with first-neighbors
hopping
yield very similar results in all cases considered.  
Our mean field results  are in agreement with the
predictions of Lieb's theorem\cite{Lieb89} regarding the total 
spin $S$ of the exact ground state of the Hubbard model in
bipartite lattices. The honeycomb lattice  of graphene is formed by two
triangular interpenetrating sublattices, $A$ and $B$. 
 Triangular nanostructures have
more atoms in one sublattice, $N_{\rm A}\neq N_{\rm B}$;
our mean field calculations
consistently predict that the total spin of the ground state is 
$2S=N_A-N_B$ and that is mainly
localized on the edges.
This could have been anticipated from Hund's rule and the appearance, in the
non-interacting model, of $N_A-N_B$ degenerate states with strictly zero energy.
Hexagonal nanostructures, for which  $N_A=N_B$, result in $S=0$ ground states
even when interactions are turned on.   A value of $S=0$ does not preclude,
however, an interesting magnetic behavior. In fact, we predict a quantum phase transition for
hexagons: Whereas small ones are paramagnetic,  large ones are compensated
ferrimagnets, both with $S=0$.

\begin{figure}
[hbt]
\includegraphics[width=3.0in]{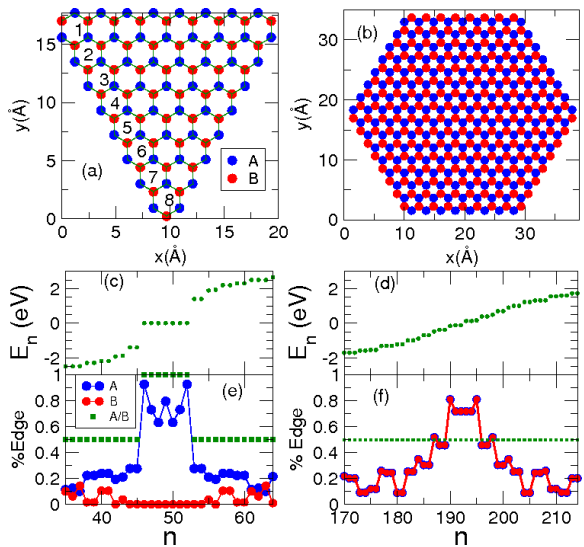}
\caption{ \label{figure1}(Color online).  
(a) and (b) Atomic structure
of the triangular and hexagonal graphene islands. 
 (c),(d) Single particle spectra for the $N=8$ triangle (left)
 and hexagon (right)  (e),(f) Sublattice resolved edge content 
 (eq. \ref{weight})  and  sublattice polarization (eq. \ref{pol}) }
\end{figure}

{\em The shape and the single-particle spectrum.-}
The different atomic structure of triangular and hexagonal graphene
nanostructures can be appreciated in  Figs. \ref{figure1}(a) and (b).
Zigzag edges are formed by atoms  that
belong to the same sublattice, $A$ or $B$.
In the case of the triangular systems
the three edges belong to the same sublattice, hereafter $A$, whereas in the
hexagon three edges are $A$-type and the other three are $B$-type.  The edge
imbalance in triangular nanostructures results in a global imbalance, so that
the total number of atoms in the sublattice  $A$ and $B$ is not the same. In
what follows we characterize the size of both triangular and hexagonal
nanostructures by the  integer number $N$ of edge atoms of the same sublattice
 along
one edge of the island (see Fig. \ref{figure1}). 
 
The structural differences between hexagonal and triangular nanostructures
reflect in their electronic properties.  
At the simplest level, we describe them 
 in the one-orbital tight-binding approximation 
 \cite{Nakada96,Fujita96,Ezawa06,Brey06,Fede06}.
The model Hamiltonian $H_0$ is totally defined by the positions of the atoms
and  the first-neighbour hopping parameter $t$, which we take equal to
2.5 eV. We set the 
on-site energies for all the carbon atoms equal to zero.
We assume that
the edge atoms are pasivated, so that there are no dangling bonds. 
In Figs. \ref{figure1}(c) and (d) we show the (low energy) spectra
corresponding to a triangle (left) and a hexagon (right) with edge size $N=8$.
If the system is charge neutral  
the relevant electronic states, corresponding to the highest
 occupied    and lowest unoccupied molecular orbitals (HOMOs and LUMOs)
 are around $E=0$. 
 The most striking difference between the spectra of the triangle and the
 hexagon is the existence of a cluster of zero-energy states in the case of the
 triangle.   A sufficient condition to have $N_Z$ states with
 strict zero energy in graphene structures is to have a  sub-lattice imbalance
 $N_Z=N_A-N_B$. In the case of graphene islands with triangle shape (see Fig.
 \ref{figure1}), the sublattice imbalance satisfies  $N_A-N_B=N-1$.
 
 In order to quantify the edge/bulk
 character of the single particle eigenstates $\phi_n(I)$,
 we  define their sub-lattice resolved
  edge content: 
  \begin{equation}
  W_{\eta}(n)=\sum_{I\in\eta,edge} |\phi_n(I)|^2
\label{weight}
  \end{equation}
  where $\eta=A,B$ and $I$ runs over the $N_{\eta}$  atoms. 
  We also define the sublattice polarization 
  \begin{equation}
  {\cal L}(n)=\sum_{I\in A} |\phi_n(I)|^2-\sum_{I\in B}|\phi_n(I)|^2
\label{pol}
    \end{equation}
 Both in the case of the triangle [Fig. \ref{figure1} (e)] and the hexagon [Fig. \ref{figure1}
 (f)] states  have a predominant edge character close to
 the Dirac point $(E=0)$, but, again, there are
 some differences. In the triangle, the zero energy states
 have a full sublattice polarization ${\cal
 L}=1$ and their edge content can reach almost 1.
 In the case of the hexagon there is a perfect $AB$ symmetry
 (${\cal L}=0.5$)  and the edge content does not go above 0.8.

{\em Electron-electron interactions.-}  
The manifold of $2N_Z$ zero energy states, including the spin,
 of the triangle  is half-filled. Electronic repulsions determine which of the
 $2^{N_Z}$ spin configurations  has the 
 the lowest energy. If Hund's rule operates in this system,
   the ground state
 of triangular graphene nanostructures (or any other sublattice unbalanced graphene
 systems, for that matter)  should  have a maximal magnetic moment  $2S=N_Z$. In contrast, 
 the single-particle spectra of hexagons features
 some dispersion, which acts against interaction induced spin polarization. 
 To put this on a quantitative basis, we have calculated the electronic 
structure using both a mean field decoupling of the one-orbital Hubbard model
and DFT calculations in a generalized gradient approximation (GGA) as
implemented in  the GAUSSIAN03 code\cite{Gaussian:03}, using an optimized
minimal basis set\cite{Cristiansen}.
 
\begin{figure}
[hbb]
\includegraphics[width=3.0in]{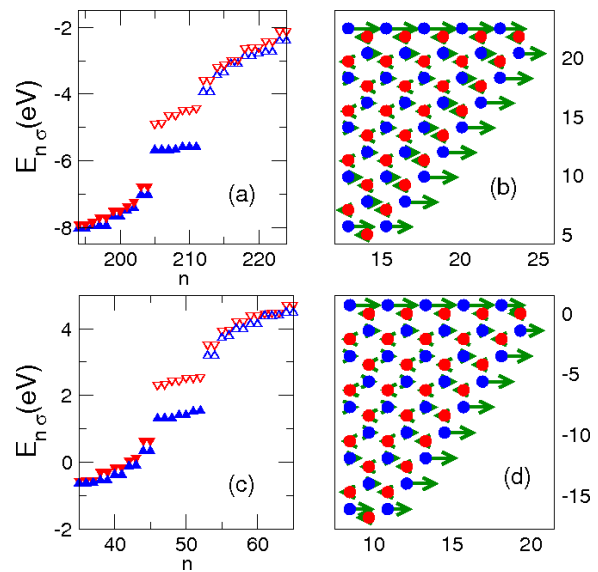}
\caption{ \label{figure2}(Color online).  Left column: self-consistent energy
spectra for a graphene triangular island with $N=8$ (fig 1.a). Closed (empty)
symbols correspond to full (empty) single particle states. 
 Right Column: local magnetization close to one of the corners of the triangle.
 Uper row: DFT calculations. Lower Row: mean field calculations with the Hubbard
 model. Magnetization arrows are plotted horizontally for the sake of clarity  }
\end{figure}
 
In the mean field approximation for the  Hubbard model
 we solve iteratively the Hamiltonian 
 \begin{equation}
 H=H_0+ U\sum_I n_{I\uparrow}\langle n_{I\downarrow}\rangle +
 n_{I\downarrow}\langle n_{I\uparrow}\rangle,
 \end{equation}
  where $H_0$ is the single
 particle Hamiltonian described above and $\langle n_{I\sigma}\rangle$ is the
 statistical expectation value of the spin-resolved density on atom $I$,
 obtained using the eigenvectors of $H$. This mean field decoupling can describe
 spontaneous symmetry breaking along a chosen axis. 
 The results shown here were obtained fixing $N_{\uparrow}$ and
 $N_{\downarrow}$, with $N_{\uparrow}+N_{\downarrow}$ equal to the number of
 carbon atoms in the structure.
  This permits to compare with DFT calculations where one typically fixes
 $N_{\uparrow}-N_{\downarrow}$.  
 A self-consistent solution of $H$ is characterized by the spin
 density: 
 $m_I\equiv \frac{n_{I,\uparrow}-n_{I,\downarrow}}{2}$,
and a single particle spectrum $\epsilon_{n,\sigma}$.
 The total spin $S=\sum_I m_I$ obviously satisfies 
 $S=\frac{N_{\uparrow}-N_{\downarrow}}{2}$.

{\em Uncompensated lattices: Triangles.-}
 In Fig. \ref{figure2} we show the spectrum and the spin density for a $N=8$
triangle.
 Upper panels correspond to DFT results with hydrogen passivation of the edge
 atoms.
  The  results in the lower panel correspond to the mean field results for the
 Hubbard model. In both cases we have verified  that the solutions with 
 $N_{\uparrow}-N_{\downarrow}=N_Z=7$ have lower ground state
 energy than solutions with
 different value of $2S$. 
 The typical energy differences are above 0.5 eV. 
   We choose the value of $U$ such that the HOMO-LUMO
 gap in the  majority spectrum is the same. In the case shown in Fig. \ref{figure2} this
corresponds to
 $U=3.85$ eV. Notice that the mean field and DFT spectrum have very similar
 structure in the neighbourhood of $E_{\rm F}$.  Interactions open a spin gap in the 
 single-particle zero-energy manifold, resulting in maximal spin polarization of
 those states. The magnetization density of both calculations is also very
 similar: The $A$ atoms on the edge are copolarized positively (right arrows)
 and their $B$ neighbour atoms are copolarized negatively. The net total spin is
 mostly sitting on the edge and the local magnetization goes to zero in the center
 of the island.  Using the same procedure as above to fix $U$, 
we find that its value decreases as the
 size of the islands increase. 
 The values of $U$ so obtained are always below
 the critical value  $U\simeq 2.2 t\simeq 5.5 eV$ above which  
 infinite graphene would
 become antiferromagnetic\cite{Fujita96,Peres04}.
 
These results indicate  that the Hubbard model 
captures the low-energy physics of graphene triangular nano-islands. One can conclude that
next-to-nearest neighbor hopping, long-range Coulomb interactions, and  
correlations,
as included in the DFT calculations,  have a minor effect on the low energy sector.
 Importantly, the basic features of the mean field
 solutions, like the structure of the spectrum, the total spin of the ground
 state and the magnetization density,
 are very robust with respect to the value of $U$. We have found very similar
 results for triangles with $N$ between 5 and 30. The solution that minimizes
 the ground state energy always satisfies 
 \begin{equation}
 2S=N_{\uparrow}-N_{\downarrow}=N_A-N_B= N_Z=N-1.
\end{equation}
 
Our mean field Hubbard model and DFT results are in agreement 
 with the Lieb theorem that states that the spin $S$ of the ground state of a
 Hubbard model in a bipartite lattice satisfies the relation
$2S=N_A-N_B$\cite{Lieb89}. If the Hubbard model with first-neighbors hopping
and constant $U$  can be
 applied to graphene-based structures of arbitrary shape, the theorem
 permits to predict the spin of the ground state by simple counting of the
 sublattice imbalance.   The fact that the number of strict zero energy
 states $N_Z$ equals to $N_A-N_B$ provides a simple picture of how the
 magnetization comes about: Spin polarization results from Hund's rule and  the
 absence of kinetic energy penalty in sublattice unbalanced graphene
 structures. 

 {\em Compensated lattices: Hexagons.-} 
 In the case of balanced structures Lieb's theorem predicts that  they
 have $S=0$. This is compatible with a locally unpolarized state, but also
with locally polarized solutions with  antiferromagnetic  correlations.
  In these cases, 
 calculations are necessary to obtain the local magnetization density. 
 In the case
 of hexagons there is a competition between the dispersion
 of the single-particle spectra and
 interactions.  Dispersion occurs because of the hybridization of states that
 otherwise would lie in a single sublattice close to the edge. These states overlap
 in the inner region  and close to the vertices 
and hybridize through hopping in $H_0$.  Smaller 
 nanostructures feature larger hybridization and are less prone to develop
 magnetic order. In the case of hexagons we expect a critical size above which 
 exchange interactions take over and the edges magnetize. This is indeed what we
 have obtained from our mean field calculations.
 
\begin{figure}
[hbt]
\includegraphics[width=2.8in]{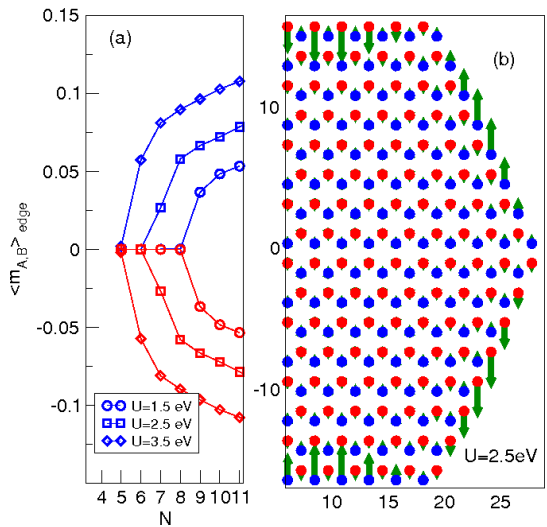}
\caption{ \label{figure3}(Color online). (a) Sublattice resolved average
magnetic moment  (eq. (\ref{m_A})) as a function of $N$ for 3 values of U.
 (b)  Magnetization density  for $U=2.5$eV and $N=8$. 
Arrows are plotted vertically  for the sake of clarity}
\end{figure}

 The local magnetization $m_I$  for the $N=8$ hexagon with $U=2.5$eV 
 is shown in the
 right panel of figure 3.
 The local magnetic moments lie mainly on the edges. 
 We quantify the formation of local moments in
 compensated structures by the sublattice resolved
 average magnetic moment on the edge atoms
  \begin{equation}
\langle m_{\eta}\rangle_{edge} =\frac{\sum'_{I\in  \eta} m_I}{3 N}
 \label{m_A}
\end{equation}
where $\eta=A,B$ and the sum runs over the   $3N$  edge atoms
of the  $\eta$ sublattice
in the hexagon. For a given value of $U$, there is a critical value
 of $N$ below
which this quantity is zero. In Fig. \ref{figure3}a we plot 
 $\langle m_{\eta}\rangle_{edge}$ as a function of $N$ for three different values
 of $U$. We always find that  
 $\langle m_{A}\rangle_{edge}=-\langle m_{B}\rangle_{edge}$.
This panel also shows how the critical size depends on $U$.
When sweeping  $U$ in a rather wide range ($U=1.5$ to $U=3.5$
 eV) the largest possible paramagnetic hexagon goes
 from 7 to 4.  We have also estimated the critical size with the help of DFT calculations
and found that the largest paramagnetic hexagon corresponds to $N=8$, 
which is consistent with the mean field Hubbard results for small $U=1.5$eV.

{\em Final remarks and conclusions.-} 
We have seen how the magnetic properties of graphene nanostructures are
intimately related to the sublattice imbalance $N_A-N_B$ in agreement with
Lieb's theorem\cite{Lieb89}. This is related to previous work on vacancies in
graphene\cite{Vozmediano05}. As a consequence of Lieb's theorem
a single vacancy results in the formation of a
local moment with $S=1/2$ and  
the sign of the spin coupling
between two single atom vacancies would depend on whether or not they belong to the same
sub-lattice \cite{kumazaki07}. 
The correlation between sublattice and sign of the exchange interaction
 is also seen in our results
   for triangular and hexagonal
nanoislands: Moments in the same sublattice couple ferromagnetically whereas 
moments in different sublattice couple  antiferromagnetically.
Indirect exchange
interaction in graphene follows the same rule \cite{RKKY}.

Nanomagnets show remanence and hysteresys because of magnetic anisotropy,
 which
originates in the spin-orbit interaction, very small in graphene. Therefore, the
{\em direction} of the spontaneous magnetization, $\vec{M}$, of graphene
nano-islands will fluctutate in the absence of an applied magnetic field. At
zero field, the detection of magnetism should rely on properties that depend on
$|\vec{M}|$, the modulus of the magnetization vector. An example of this is
the quasiparticle density of states, as probed with single electron transport in
systems with spin polarization and  without magnetic anisotropy\cite{single}. 
The controlled addition of single electrons to 
other nanomagnetic structures, like magnetic semiconductor quantum
dots\cite{magneticdots}, afford the electrical control of their magnetic
properties. This deserves further attention in the case of magnetic
graphene nano-islands.

 In conclusion, we have studied the emergence of magnetism
in graphene nano-islands with
well-defined zigzag edges.  Our DFT calculations  suggest that the
magnetic structure  of the ground state  of graphene nanoislands
can be described with a
simple Hubbard model. Ground states with finite spin $S$
appear in structures in which the number of atoms of one of the sublattices
is larger than the other, $N_A>N_B$, like triangular islands.
The single particle spectrum of
these structures features $N_Z=N_A-N_B$ states with strictly zero energy,
localized in the $A$ sublattice, which yield a magnetic ground state with 
 finite magnetic moment $S=\frac{N_A-N_B}{2}$ when interactions are included,
 both  in a mean field Hubbard model and with DFT calculations. 
 Compensated structures ($N_A=N_B$) like hexagons have $S=0$. However,
  they develop spontaneous sublattice magnetization above a critical size. 
All our results are nicely consistent with Lieb's theorem\cite{Lieb89} and complement
the theorem in the case of $S=0$ ground states.

Note added: Upon the completion of this work we have been aware of a related
work by E. Ezawa\cite{Ezawa07}, in the
single-particle approximation, and De-en Jiang {\em et al.} and 
O. Hod {\em et al.} doing
DFT calculations\cite{DFT-new}.

We acknowledge useful discussions with F. Guinea, B. Wunch, R. Miranda
 and L. Brey.
This work has been financially supported by MEC-Spain (Grants
FIS200402356  and Ramon y Cajal Program), by Generalitat Valenciana
(Accomp07-054),  by Consolider CSD2007-0010 and, in part, by   FEDER funds.

\widetext
\end{document}